# Gravity Control Propulsion: Towards a General Relativistic Approach

O. Bertolami[†]

F.G. Pedro[*]

*Instituto Superior Técnico, Departamento de Física, 1049-001 Lisboa, Portugal*

**Abstract**

Evaluation of gravity control concepts should be examined with respect to currently known physical theories. In this work we study the hypothetical conversion of gravitational potential energy into kinetic energy using the formalism of general relativity. We show that the energy involved in the process greatly exceeds the Newtonian estimate, given the nature of general relativity. We conclude that the impact of any gravity manipulation for propulsion greatly depends fundamentally on its exact definition.

[†] Associate Professor, Phone: +351-21-841-7620, Fax: +351-21-841- 9118, Email: orfeu@cosmos.ist.utl.pt
[*] Student, Email: fgpedro@fisica.ist.utl.pt

**Nomenclature**

| | |
|---|---|
| c | speed of light = $3 \times 10^8$ m.s$^{-1}$ |
| δ | inertial mass modification factor |
| ε | gravitational mass modification factor |
| ε' | gravitational field modification factor |
| G | gravitational constant = $6.67 \times 10^{-11}$ m$^3$.kg$^{-1}$.s$^{-2}$ |
| $I_{sp}$ | specific impulse |
| $v_p$ | propellant velocity |
| $G_{ik}$ | Einstein tensor |
| $T_{ik}$ | energy-momentum tensor |
| $t_{ik}$ | energy-momentum pseudo-tensor of gravitational field |
| $g_{ik}$ | metric tensor |
| g | metric determinant |
| $\Gamma^n_{lm}$ | metric connection |
| $\delta^3(\vec{r}-\vec{r}_0)$ | Dirac delta function |

## Introduction

Access to space using currently available propulsion systems is extremely limited both in distance and in time. In recent years the interest in unconventional propulsion proposals has grown in the hope that new forms of propulsion, that increase the range of spacecraft while reducing the trip time, are discovered.

Several conceptual mechanisms have been proposed to radically improve the performance of propulsion systems, such as warp-drives[1,2], transient mass fluctuations[3], antigravity[4] or gravitational shielding effects[5]. While some of these proposals are only conceptual, others such as the gravitational shielding have been tested and proved unfruitful[6]. Many of these systems are based on the effect of the manipulation of mass and gravity in a rocket's motion. The study of the impact on propulsion systems' performance of such manipulations has been previously considered and demonstrated that even if gravity could be controlled or modified it would not lead to any breakthrough in propulsion systems[7].

Other conceptual devices (e.g. the space drive) go even further, idealizing a form of propulsion without any reaction mass that would somehow manipulate space-time and matter to create propulsive forces. One of the possible energy sources for such a device, as suggested in Ref. [8], is gravitational potential energy. For concreteness we shall argue along the energy considerations as suggested in that reference. In there, it has been proposed that these systems could not be analysed using rocketry metrics and that their full potential could only be understood in terms of the energetic considerations[8]. We understand that this energy based study should be undertaken; nonetheless it should be regarded with caution, especially when trying to estimate the potential benefits of converting gravitational potential energy into kinetic energy. Furthermore, we believe that the results of any approach in the context of the Newtonian framework should be considered with great care.

In the current work we approach this space drive problem from the general relativity point of view. For this purpose we consider the energy-momentum pseudo-tensor in order to estimate the energy in a given of space-time volume. As expected, we find that the resultant energy is considerably larger than the Newtonian estimate based on the difference in the gravitational potential between two distinct points.

Before we present our computation and discuss its implications, we review some of the results presented in Ref. [7] concerning gravity manipulation based on rocketry metrics.

# Rocketry Metrics

Classical propulsion systems rely on Newton's mechanics. The foundations of these propulsion devices lie on the conservation of linear momentum in a variable mass system composed of a rocket and its propellant.

The existence of hypothetical devices capable of manipulating gravity or mass and their influence on propulsion breakthrough concepts has been studied in the Newtonian mechanics framework[7]. There have been analysed several possible manipulations:

- Inertial mass manipulation (scaling of the inertial mass: $m_i \rightarrow \delta\, m_i$),
- Gravitational mass manipulation (scaling of the gravitational mass: $m_g \rightarrow \varepsilon\, m_g$),
- Gravitational field manipulation (scaling of the gravitational coupling $G \rightarrow \varepsilon'\, G$).

It has been shown that even if achievable, these manipulations would not imply a breakthrough for propulsion, and in some cases they would have to compete with the existing technologies.

The results of these hypothetical manipulations can be summarized as follows:

| Modification | Grav. Field $\varepsilon'$ | Grav. Mass $\varepsilon$ | Inertial Mass $\delta$ | $I_{sp}$ | $v_p$ | $\Delta v$ |
|---|---|---|---|---|---|---|
| Inertial Mass | 1 | 1 | $\delta$ | $\delta^{-1/2}$ | $\delta^{-1/2}$ | $\delta^{-1/2}$ |
| Gravitational Mass | 1 | $\varepsilon$ | 1 | 1 | 1 | $\varepsilon^{1/2}$ |
| Gravitational Field | $\varepsilon'$ | 1 | 1 | 1 | 1 | $\varepsilon'^{1/2}$ |

Table 1. Summary of results of the hypothetical manipulations

For a full discussion, see Ref. [7].

# General Relativity Inspired Energy Considerations

General relativity is currently the theory that best describes gravitational phenomena having been tested in a wide variety of situations, from the solar system up to larger scales (see e.g. Ref. [10] for a review).

In the context of general relativity, gravity is interpreted as the curvature of a 4-dimensional space-time. The fundamental equations of general relativity are the Einstein's Field Equations:

$$G_{ik} = \frac{8\pi G}{c^4} T_{ik},  \tag{1}$$

where $G_{ik}$ is the Einstein tensor ($G_{ik} = R_{ik} - \frac{1}{2} g_{ik} R$) and $T_{ik}$ the energy-momentum tensor of matter.

Given that general relativity is the theory that best fits available data at solar system scale and beyond, it provides the most suitable framework to evaluate the potential breakthrough for propulsion of any system that hypothetically converts gravitational potential energy into kinetic energy.

We argue that in dealing with gravity manipulation using Newtonian mechanics is somewhat misleading given that Newton's gravity is concerned only with the dynamics along field lines. Given the nature of general relativity one should expect that any manipulation of gravity has global consequences since it would affect the space-time in its surroundings. In a previous approach it has been suggested that the amount of energy available for conversion in a trip between two points separated by a distance L in a Newtonian gravitational field is given by [8]:

$$U_N = \frac{GMm}{L} \tag{2}$$

However since general relativity describes space-time dynamics as a whole it allows for a more accurate understanding of the impact of a hypothetical gravity manipulation.

Energy conservation in a certain volume of space requires that any hypothetical gravity manipulation can convert only a fraction of the energy available in the volume element into kinetic energy and for that it must have energy resources to manipulate the space-time in question.

In order to estimate this energy, we compute the energy available in a space-time volume near a spherically symmetric mass, like the Sun. To do so it is necessary to introduce some general relativistic formalism.

The energy-momentum conservation of a system composed of matter and a gravitational field can be expressed as[11,12]:

$$\frac{\partial}{\partial x^k}(-g)(T^{ik}+t^{ik})=0, \quad (3)$$

where $T^{ik}$ is the matter energy-momentum tensor and $t^{ik}$ is the energy-momentum pseudo-tensor.

The energy-momentum pseudo-tensor $t^{ik}$ can be written as a function of the affine connection as[11,12]:

$$t^{ik} = \frac{c^4}{16\pi G} \{(2\Gamma^n_{lm}\Gamma^p_{np} - \Gamma^n_{lp}\Gamma^p_{mn} - \Gamma^n_{\ln}\Gamma^p_{mp})(g^{il}g^{km} - g^{ik}g^{lm}) +$$
$$+ g^{il}g^{mn}(\Gamma^k_{lp}\Gamma^p_{mn} + \Gamma^k_{mn}\Gamma^p_{lp} - \Gamma^k_{np}\Gamma^p_{lm} - \Gamma^k_{lm}\Gamma^p_{np}) +$$
$$+ g^{kl}g^{mn}(\Gamma^i_{lp}\Gamma^p_{mn} + \Gamma^i_{mn}\Gamma^p_{lp} - \Gamma^i_{np}\Gamma^p_{lm} - \Gamma^i_{lm}\Gamma^p_{np}) +$$
$$+ g^{lm}g^{np}(\Gamma^i_{\ln}\Gamma^k_{mp} - \Gamma^i_{lm}\Gamma^k_{np})\} \quad (4)$$

From Eq. (3) it is clear that the following quantity is conserved:

$$P^i = \frac{1}{c}\int(-g)(T^{ik}+t^{ik})dx^k, \quad (5)$$

being $P^i$ the 4-momentum of the matter plus the gravitational field.

Integrating over a hypersurface of constant time, $P^i$ can be written in the form of a 3-dimensional space integral:

$$P^i = \frac{1}{c}\int (-g)(T^{i0} + t^{i0})dV, \qquad (6)$$

To evaluate the amount of potential gravitational energy available in a certain volume of space it is necessary to find the corresponding $P^0$:

$$P^0 = \frac{1}{c}\int (-g)(T^{00} + t^{00})dV, \qquad (7)$$

Of course, the energy-momentum pseudo-tensor formalism is just an approximation to the complex problem of defining mass in general relativity without ambiguity. Nevertheless, since our computation is carried in the weak field limit, that is, in post-Newtonian limit, it is fairly accurate for our purposes.

It is now necessary to specify the space-time metric; in a weak field and low-velocity approximation, the metric can be written as:

$$g_{00} = -1 + \frac{2GM}{c^2 r} \qquad (8)$$
$$g_{11} = g_{22} = g_{33} = 1 + \frac{2GM}{c^2 r}$$

Hence, the (00) component of the energy-momentum pseudo-tensor of the gravitational field that accounts for the gravitational energy density, becomes:

$$t^{00} = \frac{7}{8\pi G}\frac{\left(\frac{\partial}{\partial r}U\right)^2}{\left(-1 + 2\frac{U}{c^2}\right)\left(1 + 2\frac{U}{c^2}\right)^3}, \qquad (9)$$

where $U \equiv \frac{GM}{r}$ is the Newtonian potential.

To calculate $P^0$ it is also necessary to know $T^{00}$, the space-drive energy density.

Assuming that the space drive is a point-like particle at rest in a position $\vec{r}_0$ relative to the centre of the spherical mass, $T^{00}$ can be then written as:

$$T^{00} = m_0 c^2 \delta^3(\vec{r} - \vec{r}_0), \qquad (10)$$

We have now an explicit formula for the energy (c $P^0$) contained within the considered volume element:

$$P^0 = \frac{1}{c}\int(-g)\left\{m_0 c^2 \delta^3(\vec{r}-\vec{r}_0) + \frac{7}{8\pi G}\frac{\left(\frac{\partial}{\partial r}U\right)^2}{\left(-1+2\frac{U}{c^2}\right)\left(1+2\frac{U}{c^2}\right)^3}\right\}dV. \quad (11)$$

In this equation there are two distinct contributions for the energy in the volume element: the first refers to the space-drive ($P^0_{sd}$), the second concerns the gravitational field through the energy-momentum pseudo-tensor ($P^0_{gf}$). It possible to write Eq. (11) as:

$$P^0 = P^0_{gf} + P^0_{sd}, \quad (12)$$

where

$$P^0_{gf} = \frac{1}{c}\int(-g)\frac{7}{8\pi G}\frac{\left(\frac{\partial}{\partial r}U\right)^2}{\left(-1+2\frac{U}{c^2}\right)\left(1+2\frac{U}{c^2}\right)^3}dV, \quad (13)$$

and

$$P^0_{sd} = \frac{1}{c}\int(-g)m_0 c^2 \delta^3(\vec{r}-\vec{r}_0)dV. \quad (14)$$

To compute these integrals is necessary to choose a particular geometrical setting. Considering that we are dealing with a device that somehow manipulates space-time in the surroundings of the spacecraft as it moves, it is hence natural to consider a cylinder as the volume element. Of course one could consider a different geometrical configuration, i.e. a conical geometry, given the radial nature of gravity. However, this would not change significantly our results. For the proposed setting, a cylinder type configuration seems more appropriate, but this is for sure not a fundamental issue. The key question is that general relativity suggests that gravity manipulation involves necessarily a volume integration.

Computing these integrals in cylindrical coordinates with with $0<r<R$, $0<\theta<2\pi$ and $R_{sun}<z<L$ (see Figure 1) we obtain:

$$P^o_{gf} = -\frac{7}{24}\frac{GM^2 R^2}{c}\left[\frac{1}{R^3_{sun}} - \frac{1}{L^3}\right] \quad (15)$$

and

$$P_{sd}{}^0 = -\frac{m_0 c^2}{c}\left[1 + 4\frac{GM}{c^2 r_0}\right], \quad (16)$$

where from the metric determinant in Eq. (14) we kept only the terms up to first order in U to compute the integral.

We point out that a computation using the higher order metric:

$$g_{00} = -1 + \frac{2GM}{c^2 r} - 2\left(\frac{2GM}{c^2 r}\right)^2 \quad (17)$$

$$g_{11} = g_{22} = g_{33} = 1 + \frac{2GM}{c^2 r}$$

would yield essentially similar results: the gravitational field component ($P^0_{gf}$) would not change at all while the space-drive component ($P^0_{sd}$) would only be different by a higher order terms in U.

We can now see that the energy available in a cylinder of radius R and length L measured form the surface of the Sun is:

$$E = cP^o = -\frac{7GM^2 R^2}{24}\left[\frac{1}{R^3_{sun}} - \frac{1}{L^3}\right] - m_0 c^2\left[1 + 4\frac{GM}{c^2 r_0}\right]. \quad (18)$$

In order to compare with the Newtonian estimate we assume that the space drive is able to manipulate a very large region of space, such that $R \approx R_{sun}$ and $L \gg R_{sun}$. It is also assumed that in this regime the dominant term in Eq. (18) is $P^o_{gf}$. Under these assumptions:

$$\frac{P^o_{gf}}{U_N} \propto \left(\frac{M_{Sun}}{m_0}\right)\left(\frac{L}{R_{Sun}}\right), \quad (19)$$

and therefore $P_{gf}^{o} \gg U$.

This result should be regarded with caution since its significance greatly depends on what we define as gravity control. Defining gravity control simply as the conversion of potential energy in a certain volume into kinetic energy then Eq. (19) tells us that there is more energy available for conversion than in the Newtonian approach. Such a result is due to the fact that in this general relativity point of view we are dealing with a certain volume of space-time instead of just two distinct points in space. On the other hand if we understand gravity manipulation as the control over a region of space-time to extract energy from it, Eq (19) represents the energy required to manipulate the considered volume of space-time. In this sense and under the specified conditions, we have shown that the required energy greatly exceeds the Newtonian energy estimate. In the former interpretation of gravity manipulation required to change space-time leads to no energy gain for spacecraft propulsion.

## Conclusions and Outlook

The possibility of manipulating gravity to extend space travelling to interstellar distances and to reduce trip time is an interesting topic from the physics and from the propulsion point of view. Evaluating the potential of any gravity manipulation concept must be carried out in the context of the currently theories of physics.

In this work we approached these issues using the general relativity framework. Through the use of the energy-momentum pseudo-tensor of the gravitational field we estimated the energy contained in a volume of space-time. Our results reveal that there is more potential energy than in the previous Newtonian estimate. However the interpretation of this result greatly depends on the exact definition of gravity manipulation. Understanding our calculation as the energy available for conversion leads to an encouraging conclusion, since the energy available is much larger than previously estimated. On the other hand, regarding this result as the energy that must be spent to control a region of space-time, leads to a radically different conclusion. From this point of view, gravity manipulation is an essentially unfruitful process for propulsion purposes.

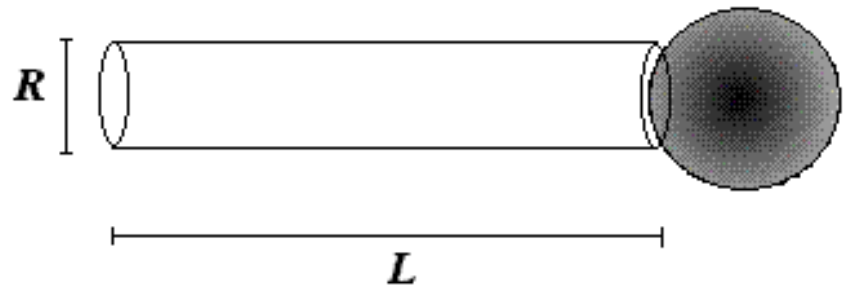

Fig.1 Volume element considered in the general relativistic computation